# HIGH REDSHIFT HCN EMISSION: DENSE STAR-FORMING MOLECULAR GAS IN IRAS F10214+4724

P. A. VANDEN BOUT
National Radio Astronomy Observatory, Charlottesville, VA 22903

P. M. SOLOMON
Department of Physics and Astronomy, SUNY, Stony Brook, NY 11794

R. J. MADDALENA
National Radio Astronomy Observatory, Green Bank, WV 24944

## ABSTRACT

Hydrogen cyanide emission in the J=1-0 transition has been detected at redshift z=2.2858 in IRAS F10214+4724 using the Green Bank Telescope[1]. This is the second detection of HCN emission at high redshift. The large HCN line luminosity in F10214 is similar to that in the Cloverleaf (z=2.6) and the ultra-luminous infrared galaxies Mrk231 and Arp220. This is also true of the ratio of HCN to CO luminosities. The ratio of far-infrared luminosity to HCN luminosity, an indicator of the star formation rate per solar mass of dense gas, follows the correlation found for normal spirals and infrared luminous starburst galaxies. F10214 clearly contains a starburst that contributes, together with its embedded quasar, to its overall infrared luminosity. A new technique for removing spectral baselines in the search for weak, broad emission lines is presented.
*Subject Headings:* galaxies:individual(F10214+4724) – galaxies:high-redshift – galaxies:starburst – galaxies:ISM – galaxies:active – radio lines:galaxies.

## 1. INTRODUCTION

This paper reports the detection of HCN (J=1-0) line emission from IRAS F10214+4724 at a redshift of z = 2.2858. The optical identification and redshift determination of this IRAS Faint Source Catalog object by Rowan-Robinson et al. (1991) established it as having a huge infrared luminosity ($\sim 10^{14}$ $L_\odot$). Subsequently, the detection of CO emission revealed the presence of a large amount ($\sim 10^{11}$ $M_\odot$) of molecular gas (Brown and Vanden Bout, 1991; Solomon et al., 1992a). Prior to this detection, high redshift (z > 2) HCN emission had been seen only from the Cloverleaf quasar (Solomon, et al. 2003). HCN emission is a more reliable signpost of dense, star-forming molecular gas than CO in that its critical density for excitation of the J=1-0 line is n $\sim 10^5$ cm$^{-3}$, whereas a density as low as n $\sim$ 300 cm$^{-3}$ can excite the CO J=1-0 line. In normal spirals, and both luminous and ultra-luminous infrared galaxies (LIRGs & ULIRGs), it has been shown that the correlation between infrared (IR) luminosity and HCN line luminosity is much tighter than that of IR with CO line luminosity (Gao and Solomon, 2004a). The star formation rate deduced from the IR luminosity scales linearly with the amount of dense molecular gas traced by HCN emission over more than three orders of magnitude in infrared luminosity. This is not the case for molecular gas traced by CO emission.

---

[1] The Green Bank Telescope is a facility of the National Radio Astronomy Observatory, operated by Associated Universities, Inc. under a Cooperative Agreement with the National Science Foundation.

Observations of HCN could be helpful in distinguishing the roles of starbursts vs. active galactic nuclei (AGNs) as the source of the high infrared luminosities observed in the known high redshift molecular line emission galaxies (HiRMLEGs). (See Hainline et al. 2004 for a current list of HiRMLEGs.) Accordingly, following our detection of HCN emission in the Cloverleaf, we have started a program of searching for HCN emission in other HiRMLEGs, starting with IRAS F10214+4724. Like the Cloverleaf, the molecular line source in F10214 is magnified by a gravitational lens. These two sources exhibit some of the strongest CO lines of all the HiRMLEGs and are likely to have the strongest HCN lines as well.

## 2. OBSERVATIONS AND RESULTS

The observations were made with the Green Bank Telescope (GBT) in February, 2004. (For a description of the GBT, see Jewell and Prestage (2004), or Prestage and Maddalena (2003).) At the redshift of F10214, the HCN (J=1-0) line is seen at 26.978 GHz. This frequency lies outside the nominal operating band of the GBT dual-beam 18-26 GHz receiver, but laboratory tests showed that the receiver exhibited good performance nonetheless. Because the receiver had not been calibrated outside of its nominal operating band, we used observations of 3C147 to determine the gain of the antenna and measure the intensities of the receiver's calibration diodes. We assumed a flux density of $1.64 \pm 0.11$ Jy for 3C147 at 27 GHz, interpolated from the 23 and 32 GHz fluxes reported by Mason et al. (1999), and Peng et al. (2000). We used existing elevation gain curves of the GBT at 43 and 20 GHz to estimate the elevation-dependent gain of the antenna at 27 GHz. We find the system gain, averaged over the elevations of our data, to be $0.75 \pm 0.08$ Jy K$^{-1}$. Zenith atmospheric opacities, as determined from measured system temperatures at various elevations, were low (~0.019 nepers) and varied little from night to night. The accuracy of our calibration is approximately 15%, almost all of which is due to the uncertainties in the assumed 3C147 flux and system gain curves. System temperatures at a typical observing elevation for the four receiver amplifiers ranged between 40 and 65 K on six observing nights of excellent winter weather.

We used the NOD observing mode, whereby each beam alternates between the source position and an off-position 3 arc-minutes away in azimuth, with one of the two beams always on the source. Scan durations were 2 minutes and each ON and OFF is saved separately for each beam and each polarization. The scans were taken as follows:

| | |
|---|---|
| Scan N | Source in beam 1, sky in beam 2; |
| Scan N+1 | Sky in beam 1, source in beam 2; |
| Scan N+2 | Source in beam 1, sky in beam 2; |
| Scan N+3 | Sky in beam 1, source in beam 2; |
| etc. | |

The spectrometer output was analyzed as follows. Each data set of 200 MHz total bandwidth divided into 8192 channels of width 24 kHz (0.27 km s$^{-1}$) each was smoothed to a velocity resolution of 17 km s$^{-1}$. For each pair of observations a normalized ON-OFF was obtained for each beam and each polarization, and their values converted from arbitrary spectrometer units to temperature units by multiplying by $T_{sys}$:

> (Scan N_beam_1 – Scan N+1_beam_1)/Scan N+1_beam_1;
> (Scan N+1_beam_2 – Scan N_beam_2)/Scan N_beam_2;
> (Scan N+2_beam_1 – Scan N+3_beam_1)/Scan_N+3_beam_1;
> (Scan N+3_beam_2 – Scan N+2_beam_2)/Scan N+2_beam_2;
> etc;

The spectrum T(v) was obtained by averaging this sequence of (ON–OFF)s with weights for successive scans of $(T_{sys})^{-2}$.

While the resulting spectrum clearly indicated the presence of a line, there was also a significant baseline curvature present. This is not surprising, since there is a two minute time delay between the ON and OFF, with resultant slight differences in the total system power from the source and the sky. The baseline is the limiting factor in detecting very weak, broad spectral lines. The standard procedure of fitting the baseline T(v) with a polynomial or sine wave and subtracting it from the spectrum can be arbitrary, and may either enhance or eliminate any true signal, particularly, from a broad spectral line.

In order to view the baseline without the presence of the line and to remove the baseline with a minimum of bias, we used the output from the dual beam system to obtain a baseline shape directly from the data. Thus, a second array B(v) is obtained from an average of (ON–ON)s and (OFF–OFF)s in the following manner:

> (Scan N_beam_1 – Scan N+2_beam_1)/Scan N+2_beam_1;
> (Scan N_beam_2 – Scan N+2_beam_2)/Scan N+2_beam_2;
> (Scan N+1_beam_1 – Scan N+3_beam_1)/Scan N+3_beam_1;
> (Scan N+1_beam_2 – Scan N+3_beam_2)/Scan N+3_beam_2;
> etc.

As for T(v), the (ON–ON)s and (OFF–OFF)s in B(v) are expressed in temperature units. The array B(v) represents the frequency dependence of the total power difference in each beam separated by a time of four minutes instead of two minutes. B(v) contains the same baseline signature as T(v) with a higher amplitude due to the longer time delay but, for weak spectral lines, without any signature of a line. We use B(v) as a template to obtain the shape and amplitude of the baseline in the T(v) data. For each polarization the average T array is fit to the average B array with

$$T(v) = a\, B(v) + b\, v + c,$$

where a is the amplitude of the B array with respect to the T array, b is a linear slope, and c is an offset. The fit excludes data in the velocity range of the possible spectral line. For F10214, the expected velocity range is known precisely from the published CO data and is only 200 km s$^{-1}$ out of a total bandpass of 2,000 km s$^{-1}$. A typical linear baseline fit would use only the two parameters b and c. All of the non-linearity, or shape, of the baseline is contained in the B(v) array, and is obtained purely from the data. Figure 1a and 1b show the data T(v) and the scaled baseline for the left and right polarizations. The HCN line is clearly present in Figure 1a and the baseline is relatively well behaved showing a curvature which can be fit with a second order polynomial. Figure 1c shows the left polarization spectrum T(v) after subtracting a second order polynomial; the HCN line is present with an intensity of about 0.0006 K.

As can be seen in Figure 1b, the right polarization baseline is complex and cannot be simply modeled by a low order polynomial. In order to make full use of the data from both polarizations the spectrum was obtained by subtracting the baseline fit from the ON–OFF data over the full range of the data:

$$T'(v) = T(v) - [a\,B(v) + b\,v + c].$$

Figure 1d shows the antenna temperature $T'(v)$ obtained by averaging both polarizations. The HCN line is clearly present and the combined signal from a total integration time on source of 42 hours has lower noise than the spectrum from the single polarization in Figure 1c. The HCN line is seen with a signal to noise ratio of $5\sigma$ at a redshift $z = 2.2858 \pm 0.0002$, essentially identical to the best redshift for F10214, from the weighted average of the CO (7-6), (6-5), and (3-2) lines: $z(CO) = 2.28581 \pm 0.00005$ (Downes and Solomon, 2004).

The observed line parameters obtained from Figure 1d are shown in Table 1, where we have used the point source sensitivity of the GBT to convert the observed line brightness temperature $T'(v)$ to flux density. We note that the $140 \pm 30$ km s$^{-1}$ width of the HCN line is marginally less than that of the CO lines. For example, the width of the CO(J=3-2) line is $220 \pm 20$ km s$^{-1}$ and that of the higher excitation CO(J=7-6) line is $210 \pm 40$ km s$^{-1}$.

## 3. DISCUSSION

The intrinsic source quantities shown in Table 2 for IRAS F10214 were calculated using the CO lens magnification of m=12 for the molecular lines and FIR radiation, and a magnification of 34 for the mid-infrared radiation (Downes et al., 1995; Downes and Solomon, 2004). We used a luminosity distance $D_L=18.3$ Gpc ($\Omega=1$ cosmology with $\Omega_m=0.3$, $\Omega_\Lambda=0.7$, and $H_0=70$ km s$^{-1}$ Mpc$^{-1}$). The line luminosity $L'$ is surface brightness times emitting area, in brightness temperature units (Solomon et al. 1997). After adjusting the luminosities from Downes and Solomon (2004) to the cosmology used here, the intrinsic infrared (10-2000μm) and far-infrared (42.5-122.5μm) luminosities, L(IR) and L(FIR), and the CO line luminosity, L'(CO 3-2), of IRAS F10214 are $9.2 \times 10^{12}$ L$_\odot$, $4.9 \times 10^{12}$ L$_\odot$, and $7.2 \times 10^{9}$ K km s$^{-1}$ pc$^2$, respectively. We have taken L'(CO 1-0) equal to $1.1 \times$ L'(CO 3-2) for both sources, following the analysis of the CO lines for the Cloverleaf of Weiß, et al. (2003). This is consistent with the limits on CO (1-0) of Tsuboi, et al. (1999) for the Cloverleaf, and of Barvainis, et al. (1995) for F10214.

IRAS F10214 has an intrinsic HCN(J=1-0) line luminosity similar to that of Mrk231 and Arp220, two prototypical ultra-luminous infrared galaxies and slightly smaller than that of the Cloverleaf. The high ratio L'(HCN 1-0)/ L'(CO 1-0) = 0.18, similar to that found in most LIRGs and ULIRGs, indicates that a large fraction of the molecular gas is dense molecular gas normally associated with giant molecular cloud cores, the sites of high mass star formation (Gao and Solomon, 2004a). In the Gao and Solomon sample of 62 galaxies with HCN observations, all galaxies with global HCN/CO luminosity ratios greater than 0.07 were found to be luminous (L(IR) > $10^{11}$ L$_\odot$) starbursts.

For galaxies in the sample of Gao and Solomon (2004a,b), the ratio L(IR)/L´(HCN 1-0) is the same for normal spirals and LIRGs, and slightly higher for ULIRGs. This indicates that the star formation rate, as indicated by L(IR), per solar mass of dense molecular gas, as indicated by L´HCN), is constant for a wide variety of galaxies including high luminosity starbursts with L(IR) > $10^{11}$ L$_\odot$. Table 2 shows that for F10214 and the Cloverleaf the ratio L(IR)/L´(HCN 1-0) is substantially higher than that for ULIRGs, LIRGs, and normal galaxies. For both sources, this is due to a large contribution to the luminosity at mid-infrared wavelengths from very hot dust heated by the quasar (and not from star formation), which clearly contributes to the overall infrared luminosity. The spectral energy distributions of both sources have been decomposed into FIR and mid-infrared components (Weiß, et al. 2003; Downes and Solomon, 2004). If we restrict the ratio to the FIR component, then L(FIR)/L´(HCN 1-0) = 2600 for F10214, only a factor of 2 higher than that of Arp220 and the average for local infrared luminous starburst galaxies. Figure 2 shows a plot of L(FIR) vs. L´(HCN 1-0) for 65 local galaxies and starbursts adapted from Gao and Solomon (2004a, 2004b). The two high z quasars lie slightly above the fit line for normal galaxies.

Using the HCN line luminosity as a quantitative star formation rate indicator (from Equation 11, Gao and Solomon, 2004a) we find a star formation rate of 340 M$_\odot$ yr$^{-1}$ if the star formation efficiency per solar mass of dense gas is the same as in local galaxies. Alternatively if all of the FIR is due to star formation, the star formation rate is about 1,000 M$_\odot$ yr$^{-1}$. It appears that about 1/5 to 1/2 of the total intrinsic infrared luminosity from F10214 can be associated with a huge starburst taking place in the dense molecular gas traced by HCN emission. These HCN observations clearly indicate the presence of a large mass of dense molecular gas in F10214 and a substantial star burst component to the FIR emission, in addition to that from dust heated by the quasar, similar to the situation in the Cloverleaf. Finally, this detection demonstrates the power of the GBT for the detection of weak, broad spectral lines. This capability will be important for future detections of high redshift molecular line emission sources.

The authors are grateful to the referee for suggestions that have improved this paper. The assistance of J. Barrett (SUNY – Stony Brook) with data analysis and figure construction is appreciated. PVB acknowledges the hospitality of the Department of Astronomy, University of Texas at Austin during this research.

Table 1. IRAS F10214 Observational Results

| | |
|---|---|
| HCN(1-0) Redshift z | $2.2858 \pm 0.0002$ |
| Peak flux density (mJy) | $0.45 \pm 0.08$ |
| S $\Delta$v (Jy km s$^{-1}$) | $0.05 \pm 0.01$ |
| Line Width $\Delta$v (km s$^{-1}$) | $140 \pm 30$ |

Table 2. Derived Intrinsic Source Properties

| | IRAS F10214 | Cloverleaf | Mrk231 | Arp220 |
|---|---|---|---|---|
| L´(HCN 1-0) (K km s$^{-1}$ pc$^2$) | $(1.9 \pm 0.3) \times 10^9$ | $(3.2 \pm 0.5) \times 10^9$ | $2 \times 10^9$ | $1.1 \times 10^9$ |
| L(IR)/L´(HCN 1-0) | $4800 \pm 1800$ | $7700 \pm 1300$ | 1600 | 1500 |
| L(FIR)/L´(HCN 1-0) | $2600 \pm 550$ | $1700 \pm 300$ | 1100 | 1300 |
| L´(HCN 1-0)/ L´(CO 1-0) | $0.18 \pm 0.04$ | $0.06 \pm 0.01$ | 0.24 | 0.12 |

Note: The source properties of the Cloverleaf, Mrk231, and Arp220 are from Solomon et al. (2003), Gao and Solomon (2004), and Solomon et al. (1992b), respectively.

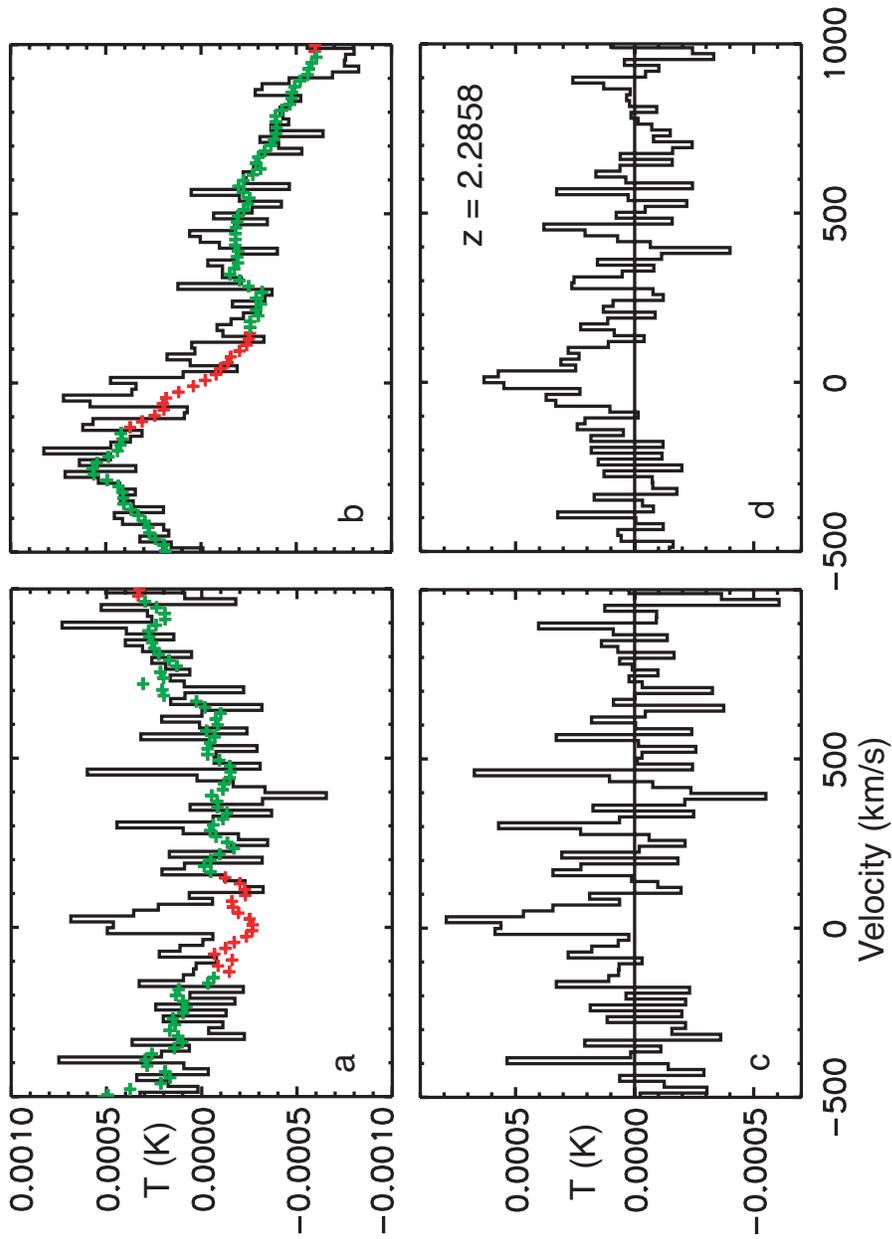

Figure 1. Illustration of the baseline removal process used in obtaining the HCN spectrum: a) the data T(v) and the scaled baseline (crosses in color) for the left polarization, where red and green indicate the velocity range excluded and included, respectively, from the fit that determines the coefficients scaling T(v) to B(v); b) the same as a) but for the right polarization; c) the left polarization data with a 2nd order polynomial fit to the baseline removed; d) the average of a) and b) after the baselines shown were removed. The improvement in signal to noise gained by the use of the new baseline removal technique is clear from the comparison of c) and d).

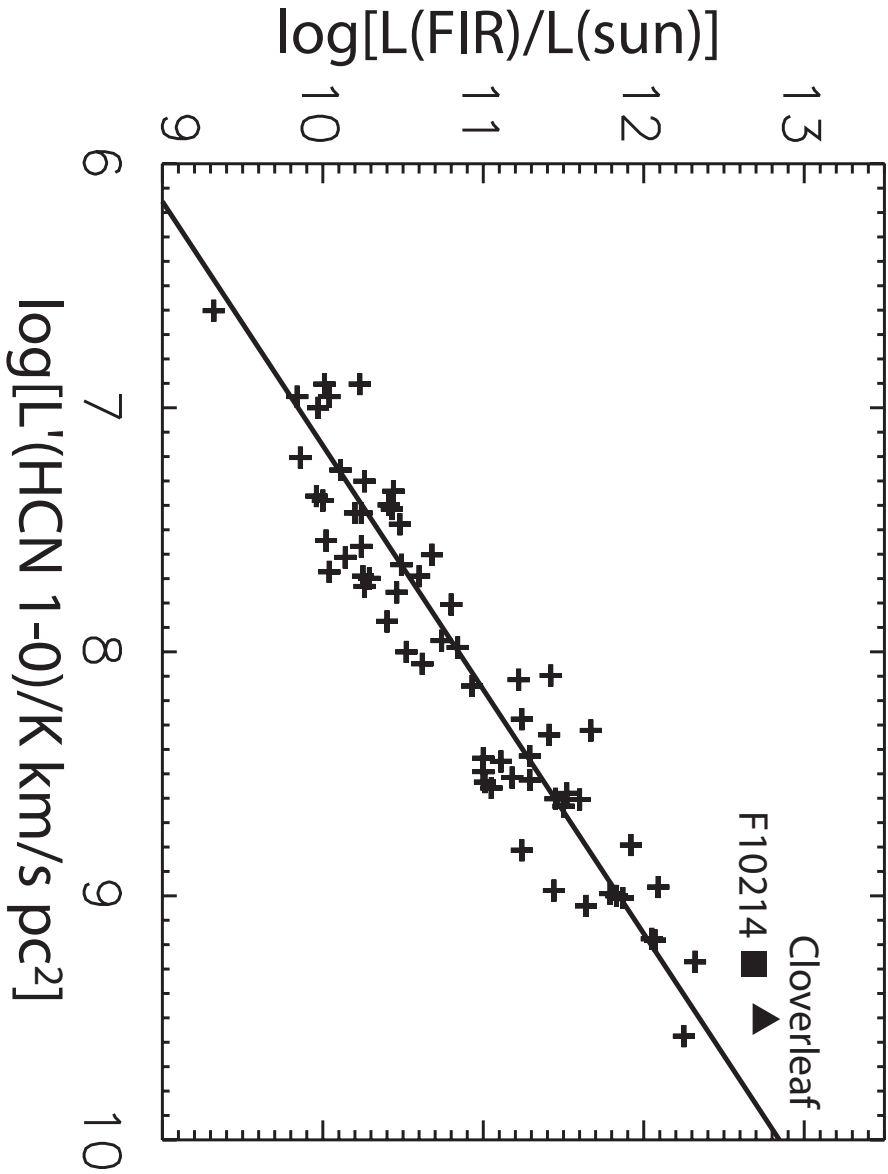

Figure 2. A plot of log L(FIR) vs. log L´(HCN 1-0) for the normal spirals, LIRGs, and ULIRGs in the sample of Gao and Solomon (2004), shown in crosses, together with the data for IRAS F10214 and the Cloverleaf. Note that this diagram plots FIR rather than IR (see text). The trend line is the linear fit to the Gao and Solomon (2004) sample with L(FIR) = 700 L´(HCN 1-0).